\documentclass[twocolumn,showpacs,preprintnumbers,amsmath,amssymb]{revtex4}
\usepackage{graphicx}% Include figure files
\usepackage{dcolumn}% Align table columns on decimal point
\usepackage{bm}% bold math
\def\mib#1{\mbox{\boldmath $#1$}}

\begin{document}

%\preprint{APS/123-QED}

\title{
Fluid dynamics and jamming in a dilatant fluid
}

\author{Hiizu Nakanishi}
\affiliation{
Department of Physics, Kyushu University 33, Fukuoka 812-8581, Japan}

\author{Namiko Mitarai}%
\affiliation{
Niels Bohr Institute, University of Copenhagen, Blegdamsvej 17, 
DK-2100 Copenhagen \O, Denmark }%
% \altaffiliation{Department of Physics, Kyushu University 33, Fukuoka 812-8581, Japan}

\date{\today, Received 30 August 2010.}% It is always \today, today,

\begin{abstract}
We present a phenomenological fluid dynamics model for a dilatant fluid,
i.e. a severe shear thickening fluid, by introducing a state variable.
The Navier-Stokes equation is coupled with the state variable field, which
evolves in response to the local shear stress as the fluid is sheared.
The viscosity is assumed to depend upon the state variable and to
diverge at a certain value due to jamming.
We demonstrate that the coupling of the fluid dynamics with the shear
thickening leads to an oscillatory instability in the shear flow.  The
model also shows a peculiar response of the fluid to a strong
external impact.
\end{abstract}


%\keywords{}

\maketitle
%-----------------------------------------------------------------------------
%\section{Introduction}

Dense mixture of starch and water is an ideal material to demonstrate
the shear thickening property of the non-Newtonian fluid. It may behave
as liquid but is immediately solidified upon sudden application of
stress, thus one can even run over the pool filled with the fluid.  It
is amazing to see that it develops protrusions sticking out of the
surface when it is subject to strong vertical
vibration\cite{Swinney-2004,Ebata-2009}, and also intriguing that the
fluid vibrates spontaneously when one simply pours it out of a
container.
A source of these unusual behaviors is the severe shear thickening; the
viscosity increases almost discontinuously by orders of magnitude at
a certain critical shear rate\cite{Bonn-2008}, 
which makes the fluid behave like a solid upon abrupt deformation.
%
%thus the fluid is
%virtually solidified upon sudden application of external stress.
%
Such severe shear thickening is often found in dense colloid suspensions
and granule-fluid
mixtures\cite{Hoffman-1972,Barnes-1989,Jaeger-2009,Laun-1991}, and they
have been sometimes called ``dilatant fluid'' due to the apparent
analogy to Reynolds dilatancy of granular media\cite{Reynolds-1885}.

%-------------------------

Despite its suggestive name, physicists have not reached the microscopic
understanding of this shear thickening property.  It was originally
proposed that the shear thickening is due to the order-disorder
transition of dispersed
particles\cite{Hoffman-1972,Hoffman-1974,Hoffman-1998,Barnes-1989}; the
viscosity increases when the fluid flow in high shear regime destroys
the layered structure that has appeared in lower shear rate regime.
Although there seem to be some situations where this mechanism was
believed to cause the shear thickening, there are other cases where
the layered structure is not observed prior to the shear
thickening\cite{Laun-1992} or no significant changes in the particle
ordering are found upon the discontinuous
thickening\cite{Wagner-2002,Wagner-2005}.  The cluster formation due to
hydrodynamic
interaction\cite{Brady-1985,Wagner-1996,Wagner-2001,MelroseBall-2004}
and the
jamming\cite{MelroseBall-2004-2,Ball-1997,Cates-1998,Bertrand-2002,
Lootens-2005,Bonn-2008,Jaeger-2009} have been proposed as plausible
mechanisms.

%---------
There are several peculiar features in the shear thickening shown by a
dilatant fluid:
(i) the thickening is so severe and instantaneous
% in response to an external stress 
that it might be used even to make a body armor to stop a
bullet\cite{Wagner-2009},
(ii) the relaxation after removal of the external stress occurs within a
few seconds, that is not very slow but not as instantaneous as in the
thickening process,
(iii) the thickened state is almost like a solid and does not allow much
elastic deformation unlike a visco-elastic material,
(iv) the viscosity shows hysteresis upon changing the shear rate\cite{Laun-1991},
(v) noisy fluctuations have been observed in response to an external
shear stress in the thickening regime\cite{Laun-1991,Lootens-2005}.

%---------
In this report, we construct a phenomenological model in the macroscopic
level and examine the fluid dynamical behavior of the
dilatant fluid.  We introduce a state variable that describes
phenomenologically an internal state of the fluid, and couple its
dynamics with the fluid dynamics.
We find that the following two aspects of the model are important: (1)
the fluid state changes in response to the shear stress, (2) the state
variable changes with the rate proportional to the shear rate.
We examine the model behavior in simple configurations and
demonstrate that the model is capable of describing the characteristic
features of hydrodynamic behavior and shows the oscillatory
instability.

%------------- Model ----------------------

\paragraph{Model:}
The model is based upon the incompressible Navier-Stokes equation with
the viscosity $\eta(\phi)$ that depends upon the state variable $\phi$.
The scaler field $\phi$, which takes a value in [0, 1], represents the
local state of the medium.  We assume two limiting states: the
low viscosity state ($\phi\sim 0$) and the high viscosity state
($\phi\sim 1$).  At $\phi=1$, the system is supposed to be jammed and
the viscosity diverges.

The state variable $\phi$ relaxes to a steady value $\phi_*$ determined
by the local shear stress $S$; $\phi_*$ is supposed to be a continuous
and monotonically increasing from 0 to $\phi_M$ as a function of $S$
with a characteristic stress $S_0$.  The limiting value $\phi_M$
represents the state of the medium in the high stress limit and should
depend upon the medium properties such as the packing fraction of the
dispersed granules.
In the following, we employ the forms
\begin{equation}
\eta(\phi) = \eta_0 \exp\left[ {A\,\phi\over 1-\phi}\right], \quad
\phi_*(S) = \phi_M {(S/S_0)^2\over 1+(S/S_0)^2}
\label{eta-Phi}
\end{equation}
with a dimensionless constant $A$.
We assume the Vogel-Fulcher type strong divergence in $\eta(\phi)$ in
oder to represent the severe thickening,
but a detailed functional form is rather arbitrary.

\begin{figure}
\centerline{\includegraphics[width=7cm]{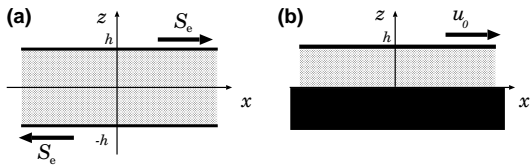}}
\caption{Simple shear flow configurations with the coordinate system.
%(a) shear flow and (b) gravitational slope flow.
}
\label{Fig-1}
\end{figure}

For the simple shear flow configuration as in Fig.\ref{Fig-1}(a) with
the flow field $\mib v$ given by $(u(z,t),0,0)$, the model is described
by the following set of equations,
\begin{eqnarray}
\rho{\partial u(z,t)\over\partial t} & = & {\partial\over\partial z} S(z,t),
\label{SF-eq-1}
\\
{\partial\phi(z,t)\over\partial t} & = &
-{1\over\tau} \Bigl(\phi(z,t) - \phi_*(S(z,t))\Bigr),
%
%\left(
%   {1\over S_0^2}\left(\eta(\phi){\partial u\over\partial z}\right)^2
%           \right),
\label{SF-eq-2}
\end{eqnarray}
with $\rho$ and $\tau$ being the medium density and the relaxation time.
The shear stress $S(z,t)$ is given by
\begin{equation}
S(z,t) = \eta(\phi)\dot\gamma, \qquad
\dot\gamma\equiv {\partial u\over \partial z},
\label{ShearStress}
\end{equation}
where the shear rate is denoted by $\dot\gamma$.

Now, we suppose the relaxation of $\phi$ toward $\phi_*$ is driven by the
shear deformation, thus the relaxation rate $1/\tau$ is not constant
but proportional to the absolute value of the shear rate, i.e.
\begin{equation}
   {1/\tau} ={|\dot\gamma|/ r}
\label{tau}
\end{equation}
with a dimensionless parameter $r$.
{
Note that, with this form of relaxation, the state variable $\phi$ does
not exceed 1 even when $\phi_M> 1$ because
$\dot\gamma=0$ at $\phi=1$ due to the diverging viscosity.
}

This particular form of relaxation is employed in order to represent the
athermal relaxation driven by local deformation.  The system responds in
accordance with the deformation rate, and does not change the state when
it does not deform.  This is natural if we consider the state change is
caused by local configurational changes of the dispersed granules in
the situation where Brownian motion does not play an important role.

%--------------------
%---------------------------------------------------------------
\paragraph{Steady shear flow:}
First, we examine the model behavior in the steady shear flow
configuration with a fixed external stress $S_{\rm e}$.  The boundary
condition is then given by
\begin{equation}
S_{\rm e}  =   S(z,t)\Bigr|_{z=\pm h} .
\label{SF-BC}
\end{equation}
The steady state solution for Eqs.(\ref{SF-eq-1}) and (\ref{SF-eq-2})
with this boundary condition is obtained easily as
\begin{equation}
\phi = \phi_*(S_{\rm e}), \quad
\dot\gamma = {S_{\rm e}\over\eta(\phi_*)} \equiv \dot\gamma_*(S_{\rm e}).
\label{SF-SS}
\end{equation}
%--------------------
\begin{figure}
%\centerline{\includegraphics[width=8cm]{./Shear-stress-shear_rate.eps}}
\centerline{\includegraphics[width=8cm]{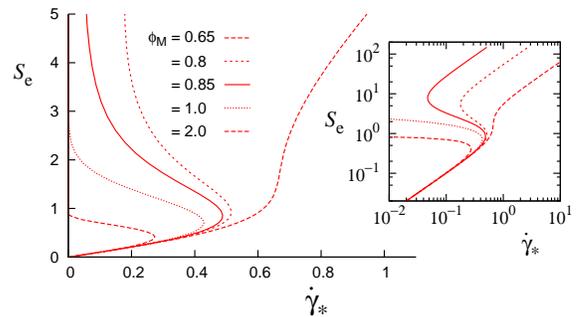}}
\caption{
The stress-shear rate relation for the viscosity given by
Eq.(\ref{eta-Phi}) with  $A=1$.
The inset shows the same plots in the logarithmic scale.
We employ the unit system where $\eta_0=S_0=\rho=1$.
}
\label{SF-gamma-S}
\end{figure} 
%--------------------

The stress-shear rate curves are plotted in Fig.\ref{SF-gamma-S} for
$\phi_*$ and $\eta(\phi)$ given by Eq.(\ref{eta-Phi}) with $A=1$ for
various values of $\phi_M$.
In the logarithmic scale, the straight line with the slope 1 corresponds
to the linear stress-shear rate relation with a constant viscosity.  
In the curve for $\phi_M=0.8$, one can see the two regimes: the low
viscosity regime and the high viscosity regime.
Between the two regimes, there is an unstable region.
From this stress-shear rate curve, we expect there should be hysteresis
upon changing the shear rate $\dot\gamma$ with discontinuous jumps between
the two branches. The jumps correspond to the discontinuous
change of viscosity, i.e. the abrupt shear thickening.
For the case $\phi_M\ge 1$, the curves do not have an upper linear branch
because the viscosity diverges and the fluid is solidified.

%---------------------------------------------------------------
\paragraph{Oscillation in the shear flow:}
If the external stress $S_{\rm e}$ is kept at a value in the unstable
branch, where
\begin{equation}
{\partial\dot\gamma_*\over\partial S_{\rm e}} < 0,
\label{unstable}
\end{equation}
i.e. the shear rate decreases as the stress increases, the steady flow
may become unstable.  The linear stability analysis within the solution
of the form $(u(z,t),0,0)$ shows that the mode whose wave number $k$ in
the $z$-axis satisfies
\begin{equation}
0 < k^2 < \rho {\dot\gamma_*\over r}\left[-{\partial\dot\gamma_*\over\partial
				 S_{\rm e}}\right] \equiv k_{\rm c}^2
\end{equation}
grows and the threshold mode $k_{\rm c}$ oscillates at 
the finite frequency
\begin{equation}
\omega_{\rm c} = \sqrt{{S_{\rm e}\over\rho\, r}}\; k_{\rm c}.
\end{equation}
%Here, the suffix $*$ denotes the steady state value.
%
{
Since the smallest
possible wave number $k_{\rm min}$ is given by $k_{\rm min}=2\pi/4h$, we
expect the oscillatory flow appears for $k_{\rm min}<k_{\rm c}$ as the
system width increases if the external stress is in the unstable branch.}

%------------------
\begin{figure}
\centerline{
\includegraphics[width=7cm]{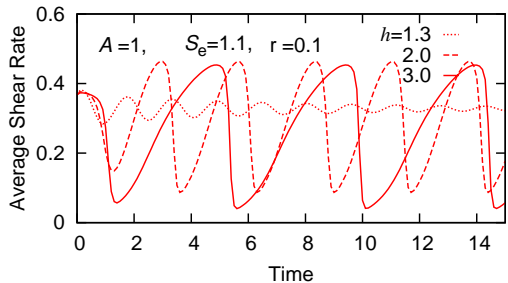}
}
\caption{Oscillation of the average shear rate $u(h)/h$ in shear flow for
 $h=$1.3, 2, and 3 with $A=1$, $\phi_M=1$, $r=0.1$, and $S_{\rm e}=1.1$.
The initial state
is chosen as the steady solution (\ref{SF-SS}) with $S_{\rm
e}=1.0$ at $t=0$.}
\label{SF-SurfV}
\end{figure}
%------------------

The oscillatory behavior of the shear flow in the unstable regime can be
seen by numerically integrating Eqs.(\ref{SF-eq-1}) and (\ref{SF-eq-2})
with (\ref{SF-BC}).
In Fig.\ref{SF-SurfV}, the average shear rate $u(h)/h$ is plotted as a
function of time for some values of the system width $2h$ for
$\phi_M=1$, $A=1$, and $r=0.1$ with the external shear stress $S_{\rm
e}=1.1$ in the unstable regime.  The initial state is chosen as the
steady solution (\ref{SF-SS}) with $S_{\rm e}=1$.
For the case of $h=1.3$, one can see the overdamped oscillation that
converges to the steady state, but for the larger values of $h$, the
system undergoes the oscillatory transition and the surface velocity
oscillates with an saw-tooth like wave profile, namely, the
gradual increase of velocity followed by a sudden drop.

\begin{figure}
\centerline{
\includegraphics[width=8cm]{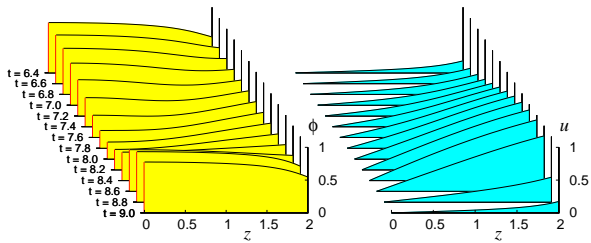}
}
\caption{Spatial variation of shear flow oscillation 
for $A=1$, $\phi_M=1$,  $r=0.1$, $S_{\rm e}=1.1$, and $h=2$.
}
\label{SF-t-develop}
\end{figure}

The time development of the spatial variations for the state variable
$\phi$ and the velocity $u$ are plotted in Fig.\ref{SF-t-develop} for
$h=2$.  Only the positive halves ($z>0$) of the symmetric solutions are
shown.
One can see the high shear rate region gradually extends towards the
center as $\phi$ relaxes, then the velocity drops suddenly to a very
small value when the shear rate exceeds a certain value and $\phi$
increases rapidly.  This sudden drop of velocity is due to the abrupt
thickening of the fluid caused by the high shear stress.  The resulting
low shear rate eventually leads to the low shear stress, which in turn
leads to small $\phi$ with low viscosity, then the shear rate starts
increasing again.

%------------------
%------------------

%---------------------------------------------------------------
\paragraph{Response to impact:}
The athermal relaxation (\ref{tau}) gives the system an interesting
feature in responding to an external impact.  We consider the simple
shear flow but assume the fixed lower boundary at $z=0$ and the upper
boundary at $z=h$ is forced to move at the velocity $u_0$ at $t=0$ (Fig.\ref{Fig-1}(b)).
Then, the motion of the upper wall is given by the equation,
\begin{equation}
m{du(h,t)\over dt} 
= -\eta(\phi)\left.{\partial u\over\partial z}\right|_{z=h},
\label{Imp-eq}
\end{equation}
where $m$ is the mass of the upper wall per unit length.

\begin{figure}
%\centerline{\includegraphics[width=8cm]{./Impulse_Disp_Jamm.eps}}
\centerline{\includegraphics[width=8cm]{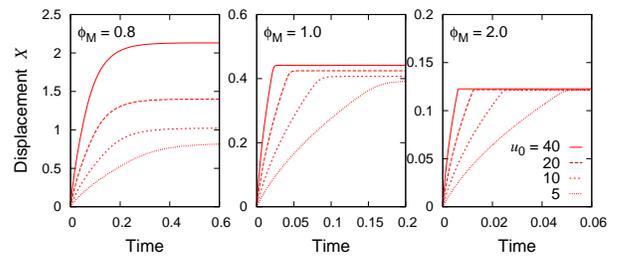}}
\caption{
The time dependence of the displacement $X$ after the impact with the
 initial speed $u_0=$40, 20, 10, and 5 for the system with $\phi_M=0.8$,
 1, and 2.
The other parameters are $h=2$, $A=1$, $r=0.1$, and $m=1$.
}
\label{Impact}
\end{figure}

Fig.\ref{Impact} shows the displacement $X$ of the upper wall,
\begin{equation}
X(t)  = \int_0^t u(h,t') dt',
\end{equation}
for the three cases, $\phi_M=$ 0.8, 1, and 2, for various initial speeds.
%
%Note that even for the case $\phi_M=2$, the actual value of $\phi$ does
%not exceed 1.  
%
The wall decelerates rapidly as the fluid thickens due to the
imposed stress, and eventually stops.  In the case of $\phi_M=0.8$, the
total distance that the wall moves before it stops increases as the
initial speed $u_0$.
On the other hand, in the case of $\phi_M=2$, the total displacement
does not depend on $u_0$.  This is because the fluid jammed at a certain
strain as it deforms, and cannot deform further.

%------------------
%------------------

%-----------------------------------------------------------------

%\section{Discussions}
%\vskip 1ex
\paragraph{Concluding remarks:}
%
% Let us discuss some of the points of the present model.
%---------
There are a couple of related models: the soft glassy rheology(SGR)
model\cite{Head-2001} and the schematic mode coupling
theory(MCT)\cite{Holmes-2005}.  In the SGR model, SGR is extended to
describe the shear thickening by introducing the stress dependent
effective temperature, which may be compared with the inverse of the
state variable $\phi$ of our model.  In the schematic MCT, the jamming
transition has been examined by means of MCT incorporating the effects
of shear schematically. Both are semi-empirical but deal with
microscopic processes and produce the similar flow curves as
Fig.\ref{SF-gamma-S}. On the other hand, the present model is
phenomenological one only for macroscopic description to study the
interplay between the fluid dynamics and the shear thickening, which is
embedded in Eqs.(\ref{eta-Phi}).

%---------

\begin{figure}
%\centerline{\includegraphics[width=6cm]{./Configuration-s-mod.eps}}
\centerline{\includegraphics[width=6cm]{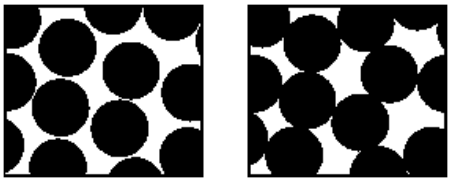}}
\caption{Schematic illustration of configurations for low stress (left)
 and high stress(right) regime.}
\label{State}
\end{figure}

Although our model is phenomenological, the microscopic picture we may
have for a dense mixture of granules and fluid is the following; in the
low stress regime ($\phi\sim 0$), the particles are dispersed thus the
medium can deform easily due to the lubrication, whereas, in the high
stress regime ($\phi\sim 1$), the particles form contacts and force
chains to support the stress, and eventually get jammed when the packing
fraction is large enough (Fig.\ref{State}).
With this picture, it is natural to assume that the time scale for the
state change is set by the shear rate in the case where the athermal
deformation drives the configurational changes, 
%
%Thus,
and then the parameters $S_0$ and $r$ correspond to the stress and the
strain, respectively, around which the neighboring granules start
touching each other.
In the case where the thermal relaxation takes place, the time scale may
be given by
\begin{equation}
{1\over\tau} = 
  {|\dot\gamma|\over r} + {1\over\tau_{\rm th}}\, ,
%  \sqrt{\left({\dot\gamma\over r}\right)^2 + {1\over\tau_{\rm th}^2}}\, ,
\end{equation}
where $\tau_{\rm th}$ is the time scale that the thermal agitation
dislocates the contacts.

%-----------

%--------------------------------------------------------------------

We have assumed that the state is determined by the shear stress, but it is
instructive to consider what would happen if the state is determined by the
shear rate.  If $\phi_*$ is a function of $\dot\gamma$ instead
of Eq.(\ref{eta-Phi}), then the stress-shear rate curve becomes
monotonically increasing without an unstable branch, thus there should
be no discontinuity and hysteresis, thus no instability that leads to
the oscillation.
This suggests that experimentally observed discontinuous
transition\cite{Bonn-2008} and the hysteresis\cite{Laun-1991} can be
attributed to the ``shear-stress thickening'' property of the medium.
%
%It should be also pointed out that the similar stress-shear rate curve
%as in Fig.\ref{SF-gamma-S} has been obtained by the mode coupling theory
%based upon the picture of the stress induced jamming\cite{Holmes-2005}.
%
%---------------
%
The direct consequence of the shear-stress thickening  may be
seen in the gravitational slope flow, where the flux does not increases
monotonically as the inclination angle because the fluid becomes more
viscous under larger shear stress, thus it may flow slower as the slope
is inclined steeper.

%---------------

%------------------------

A peculiar result of the present model is the oscillatory instability in
the shear flow. Superficially, this might look like a stick-slip motion,
but the physical origin is different; the stick-slip motion is
caused by the slip weakening friction when the system is driven at a
constant speed through an elastic device.  If the present system is
driven at a constant velocity, the system settles in one of the stable
states for a given $\dot\gamma$.
The oscillatory behavior in the present system appears under the drive
with a constant shear stress, and it is a result of the coupling between
the internal dynamics and the fluid dynamics.  In this sense, it is also
different from the oscillation in the SGR model, where the fluid
dynamics is not considered.
We found that the oscillatory flow also appears in the
gravitational slope flow.

Regarding the oscillations in experiments, one can easily notice the
vibration around 10 Hz when pouring the cornstarch-water mixture out of
a cup.  In the literature, the pronounced fluctuations in the
deformation rate have been reported in the shear stress controlled
experiment near the critical shear rate\cite{Laun-1991,Lootens-2005}
and the pressure driven flow through microchannels\cite{Isa-2009}, but
not many experiments have been reported yet.
%
%On the other hand, in the case of shear thinning system, much clearer
%oscillation have been reported in the liquid crystal
%systems\cite{Wunenburger-2001} along with the discontinuous transition
%and hysteresis\cite{Bonn-1998,Volkova-1999}, which may be explained by
%the present framework of the phenomenology.

%Chaotic dynamics has been observed in dilute aqueous solutions of a
%surfactant.  The experiment was done under the constant shear rate in
%the shear thickening regime\cite{Sood-2000,Sood-2001}, and was
%interpreted as the stick-slip transition between the two states of the
%fluid structure, thus it is not clear whether it is physically related
%to the present instability, but the chaotic dynamics can be seen in the
%present model if we set the relaxation time $\tau$ constant either in
%the case of the large system width or the short relaxation time.

This work was supported by KAKENHI(21540418).

%----------------------- Acknowledgment  ----------------

%---------------------------------------------------------
%\bibliography{references}

%---------------------------------------------------------
\end {document}